\documentclass{emulateapj}
\usepackage{natbib}
\usepackage{bm}
\usepackage{graphicx}
\usepackage{epstopdf}
\usepackage[english]{babel}
\usepackage[colorlinks,linkcolor={blue},citecolor={blue},urlcolor={blue}]{hyperref}
\newcommand{\mpchi}{\,h^{-1}{\rm {Mpc}}}

\newcommand{\msun}{\,h^{-1}{M_{\sun}}}
\shorttitle{Modeling Three-point correlation function} \shortauthors{H. Guo et al.}
\begin{document}
\title{Galaxy Three-Point Correlation Functions and Halo/Subhalo Models}

\author{Hong Guo\altaffilmark{1}, Zheng Zheng\altaffilmark{2}, Peter
S. Behroozi\altaffilmark{3,4}, Idit Zehavi\altaffilmark{5}, Johan Comparat\altaffilmark{6,7}, Ginevra Favole\altaffilmark{6,8}, Stefan Gottl\"ober\altaffilmark{9}, Anatoly Klypin\altaffilmark{10}, Francisco Prada\altaffilmark{6,8,11}, Sergio A. Rodr\'{i}guez-Torres\altaffilmark{6,7,8,12}, David H. Weinberg\altaffilmark{13,14}, and Gustavo Yepes\altaffilmark{7}}

\altaffiltext{1}{Key Laboratory for Research in Galaxies and Cosmology, Shanghai Astronomical Observatory, Shanghai 200030, China; guohong@shao.ac.cn}
\altaffiltext{2}{Department of Physics and Astronomy, University of Utah, UT 84112, USA}
\altaffiltext{3}{Astronomy and Physics Departments and Theoretical Astrophysics Center, University of California, Berkeley, CA 94720, USA}
\altaffiltext{4}{Hubble Fellow}
\altaffiltext{5}{Department of Astronomy, Case Western Reserve University, OH 44106, USA}
\altaffiltext{6}{Instituto de F\'{\i}sica Te\'orica, (UAM/CSIC), Universidad Aut\'onoma de Madrid,  Cantoblanco, E-28049 Madrid, Spain}
\altaffiltext{7}{Departamento de F{\'i}sica Te{\'o}rica,  Universidad Aut{\'o}noma de Madrid, Cantoblanco, 28049, Madrid, Spain}
\altaffiltext{8}{Campus of International Excellence UAM+CSIC, Cantoblanco, E-28049 Madrid, Spain}
\altaffiltext{9}{Leibniz-Institut fur Astrophysik (AIP), An der Sternwarte 16, D-14482 Potsdam, Germany}
\altaffiltext{10}{Astronomy Department, New Mexico State University, MSC 4500, PO Box 30001, Las Cruces, NM, 880003-8001, USA}
\altaffiltext{11}{Instituto de Astrof\'{\i}sica de Andaluc\'{\i}a (CSIC), Glorieta de
la Astronom\'{\i}a, E-18080 Granada, Spain}
\altaffiltext{12}{Campus de Excelencia Internacional UAM/CSIC Scholar}
\altaffiltext{13}{Department of Astronomy, Ohio State University, Columbus, OH 43210, USA}
\altaffiltext{14}{Center for Cosmology and Astro-Particle Physics, Ohio State
University, Columbus, OH 43210, USA}

\begin{abstract}
We present the measurements of the luminosity-dependent redshift-space three-point correlation functions (3PCFs) for the Sloan Digital Sky Survey (SDSS) DR7 Main galaxy sample. We compare the 3PCF measurements to the predictions from three different halo and subhalo models. One is the halo occupation distribution (HOD) model and the other two are extensions of the subhalo abundance matching (SHAM) model by allowing the central and satellite galaxies to have different occupation distributions in the host halos and subhalos. Parameters in all the models are chosen to best describe the projected and redshift-space two-point correlation functions (2PCFs) of the same set of galaxies. All three model predictions agree well with the 3PCF measurements for the most luminous galaxy sample, while the HOD model better performs in matching the 3PCFs of fainter samples (with luminosity threshold below $L^*$), which is similar in trend to the case of fitting the 2PCFs. The decomposition of the model 3PCFs into contributions from different types of galaxy triplets shows that on small scales the dependence of the 3PCFs on triangle shape is driven by nonlinear redshift-space distortion (and not by the intrinsic halo shape) while on large scales it reflects the filamentary structure. The decomposition also reveals more detailed differences in the three models, which are related to the radial distribution, the mean occupation function, and the velocity distribution of satellite galaxies inside halos. The results suggest that galaxy 3PCFs can further help constrain the above galaxy-halo relation and test theoretical models. 
\end{abstract}


\section{Introduction}
The spatial distribution of matter in the universe can be fully described by the two-point correlation function (2PCF) or its Fourier-space counterpart, the power spectrum, if the matter density field were completely Gaussian. However, the nonlinear gravitational evolution and structure formation naturally generates non-Gaussianity \citep[see e.g.,][]{Bernardeau02}, which produces a non-zero three-point correlation function (3PCF) even for a primordial Gaussian density field. As a biased tracer to the matter density field, galaxies also typically have non-zero 3PCF. The galaxy 3PCF describes the probability of finding galaxy triplets in the universe. It is usually used in measuring the linear and first-order nonlinear galaxy bias factors, which helps constrain cosmology by breaking the degeneracy between the linear galaxy bias and the amplitude of the matter density fluctuations (usually denoted as $\sigma_8$, the r.m.s. fluctuation of the matter density within spheres of radius $8\mpchi$; \citealt{Gaztanaga94,Jing04,Gaztanaga05,Zheng04,Pan05,Sefusatti05,Guo09b,McBride11b,Marin13,Guo14a,Moresco14}). There are also efforts of measuring the Baryon Acoustic Oscillation (BAO) features in the 3PCF \citep{Gaztanaga09,Slepian15a}.  

The galaxy 3PCF aids in learning about the galaxy-halo connection within the halo occupation distribution (HOD) framework \citep[e.g.,][]{Jing98,Peacock00,Scoccimarro01,Berlind02,Zheng05,Zheng09,Guo14b}. The amplitude of the 3PCF tightens the galaxy occupation number as a function of halo mass \citep[e.g.,][]{Kulkarni07,Smith08,Guo15a}. The shape of the 3PCF potentially teaches us about the distribution of galaxies inside dark matter halos. The purpose of this paper is to use the measured 3PCFs of galaxies to test halo-based models.

By accounting for the galaxy peculiar velocity distribution (velocity bias) inside halos, \cite{Guo15a} found that the redshift-space galaxy 3PCF is successfully explained within the HOD framework. One other popular model for inferring the galaxy-halo connection uses the subhalo abundance matching \citep[SHAM;][]{Kravtsov04,Conroy06,Vale06,Wang07,Behroozi10,Guo10,Moster10,Nuza13,Rodriguez-Puebla13,Sawala15}, by populating central and satellite galaxies into distinct halos and subhalos identified in high-resolution $N$-body simulations. \cite{Guo16} (hereafter G16) presented an extension to the SHAM model, dubbed as SCAM (subhalo clustering and abundance matching), which allows central and satellite galaxies to have different galaxy-halo relations and matches (fitting) both galaxy abundance and clustering. By comparing the modeling results for galaxy projected and redshift-space 2PCFs with the HOD, SHAM, and SCAM model, G16 concluded that the HOD and SCAM models generally produce better fits to the observed galaxy 2PCFs than the SHAM model, with the HOD model having slightly lower $\chi^2$ values. The main difference between the HOD and SCAM models lie in the spatial distribution of satellite galaxies inside halos. The HOD model used in G16 assigns satellites to randomly selected matter particles and the SCAM model put satellites in the centers of subhalos. The radial distribution of subhalos in the host halo is found to be significantly shallower than that of the dark matter \cite[see e.g.,][]{Gao04,Pujol14}. In reality, the spatial distribution of satellites may deviate from that of subhalos (e.g., caused by baryonic processes), and there is evidence that satellites follow more closely to the dark matter distribution inside halos \citep[e.g.,][]{Wang14}.

In this paper, we intend to take advantage of the added information in the galaxy 3PCF to see whether we can further distinguish the different halo models. We directly use the bestfit HOD and SCAM models from G16 that jointly fits both the projected and redshift-space 2PCFs to calculate the redshift-space 3PCF and compare with the observation. In Section~\ref{sec:data}, we briefly describe the galaxy sample, the 3PCF measurements, and the modelling of the 3PCF. We present the modelling results in Section~\ref{sec:results} and conclude in Section~\ref{sec:discussion}. 

\section{Measurements and Models} \label{sec:data}
\begin{table}
	\caption{Samples of different luminosity thresholds} \label{tab:sample}
	\begin{tabular}{@{}lrrrr@{}}
		\hline
		Sample  & $z_{\rm max}$ & $N_{\rm gal}$ & $n_g (h^{3}{\rm {Mpc}}^{-3})$  & Volume ($h^{-3}{\rm {Mpc}}^3$)\\
		\hline
		$M_r<-19$  & 0.064 & 72484 & $15.66\times10^{-3}$ & $  4.87\times10^6$\\
		$M_r<-20$  & 0.106 & 131623 & $ 6.37\times10^{-3}$ & $ 22.00\times10^6$\\
		$M_r<-21$  & 0.159 & 76802 & $ 1.16\times10^{-3}$ & $ 71.74\times10^6$\\
		\hline
	\end{tabular}
	
	\medskip
	The minimum redshift of all the luminosity threshold samples is $z_{\rm min}=0.02$. The maximum redshift $z_{\rm max}$, the total number of galaxies $N_{\rm gal}$, the mean number density $n_g$ and the volume of each sample are also displayed.
\end{table}
We use the same volume-limited luminosity threshold galaxy samples as in \cite{Guo15b}, selected from the Sloan Digital Sky Survey (SDSS; \citealt{York00}) Data Release 7 (DR7; \citealt{Abazajian09}) Main galaxy sample in the redshift range of $0.02<z<0.2$. For the analysis in this paper, we only focus on the three luminosity threshold samples that cover both faint and luminous galaxies, with $r$-band absolute magnitude as $M_r<-19$, $M_r<-20$, and $M_r<-21$. We display the sample information in Table~\ref{tab:sample} and refer the readers to Figure 1 of \cite{Guo15b} for more details. 

Following \cite{Guo15a}, we measure the redshift-space galaxy 3PCF, $\zeta(r_1,r_2,\theta)$, on small and intermediate scales using the estimator of \cite{Szapudi98}, where triangles from galaxy triplets are parametrized as $(r_1,r_2,\theta)$, with $r_1$ and $r_2$ ($r_2\ge r_1$) the lengths of two sides and $\theta$ the angle between the two sides. We use logarithmic binning schemes for $r_1$ and $r_2$ and linear bins for $\theta$, with $\Delta\log r_1=\Delta\log r_2=0.2$, and $\Delta\theta=0.1\pi$.

For the model 3PCFs, we use the measurements from the mock galaxy catalogs generated based on the bestfit HOD and SCAM models of G16 and halos/subhalos identified in high-resolution $N$-body simulations. The method adopted for the HOD and SCAM modeling of 2PCFs in G16 follows \citet{Zheng16}, which is based on $N$-body simulations and hence much more accurate than analytic models. It is equivalent to populating halos or subhalos in the simulation given a set of HOD/SCAM parameters and using the measured 2PCFs from the mock catalog as the model prediction. Since the method tabulates all halo distribution functions and clustering properties relevant for galaxy 2PCFs, it is also fast in calculating the model 2PCFs and enables an efficient exploration of the parameter space. To be fully consistent, here we use the same simulations as in the 2PCF modeling of G16.

For the two luminous galaxy samples of $M_r<-20$ and $M_r<-21$, we use the MultiDark simulation of Planck cosmology (MDPL) \citep{Klypin16}, with the cosmological parameters of $\Omega_m=0.307$, $\Omega_b=0.048$, $h=0.678$, $n_s=0.96$, and $\sigma_8=0.823$. The simulation has a volume of 1\,$h^{-3}$\,Gpc$^3$ (comoving) and a mass resolution of $1.51\times10^9\msun$. For the sample of $M_r<-19$, in order to resolve the host halos and subhalos, we use a higher-resolution simulation with the same cosmology as the MDPL but with a volume of $0.4^3\,h^{-3}$\,Gpc$^3$ and a mass resolution of $9.6\times10^7\msun$ (referred to as the SMDPL in \citealt{Klypin16}). The host halos and subhalos in the simulations are identified with the Rockstar phase-space halo finder \citep{Behroozi13}, which identifies the spherical halos from the density peaks in the phase space.

To construct the redshift-space mock galaxy catalogs, we populate host halos and subhalos in the simulations with galaxies according to the HOD and SCAM models in G16 that best describe the projected and redshift-space 2PCFs. In the models, galaxy velocity bias (the difference between galaxy and dark matter velocity distribution inside halos) is found to be required to fit the redshift-space clustering of the SDSS Main sample galaxies. The central galaxy velocity bias $\alpha_{\rm c}$ (satellite velocity bias $\alpha_{\rm s}$) is parameterized as the ratio of the dispersion of the central (satellite) galaxy motion to that of the dark matter particles in the frame defined by the halo bulk velocity (see Equations~6 and~7 in \citealt{Guo15b}).

In the HOD model, a central galaxy is put at its host halo center (defined as the position of the potential minimum), with a velocity offset with respect to the halo bulk velocity characterized by the bestfit central galaxy velocity bias $\alpha_{\rm c}$. The satellite galaxies are populated by adopting the positions of the randomly-selected dark matter particles. The velocity of a satellite galaxy is assigned through scaling the velocity of the dark matter particle relative to the halo bulk velocity by a factor of the satellite galaxy velocity bias $\alpha_{\rm s}$. In the SCAM model, the central galaxies are populated in the same way as in the HOD model. The satellite galaxies have the positions of the subhalo centers, and the velocities of subhalos relative to halo bulk velocity are scaled according to the satellite velocity bias $\alpha_{\rm s}$.

For the SCAM models, to connect the central galaxies to the host halos and the satellite galaxies to the subhalos, we consider two models with maximum circular velocities as the properties for the host halos and subhalos, i.e. the $V_{\rm acc}$ and $V_{\rm peak}$ models in G16. Both were shown to provide reasonable fits to the projected and redshift-space 2PCFs. In the $V_{\rm acc}$ model, the maximum circular velocity is used for the host halos, and the maximum circular velocity at the time of accreting to host halos is adopted for the subhalos. In the $V_{\rm peak}$ model, the peak circular velocity over the entire merger history is used for both the host halos and subhalos.

\section{Results} \label{sec:results}
\begin{figure*}
\includegraphics[width=1.0\textwidth]{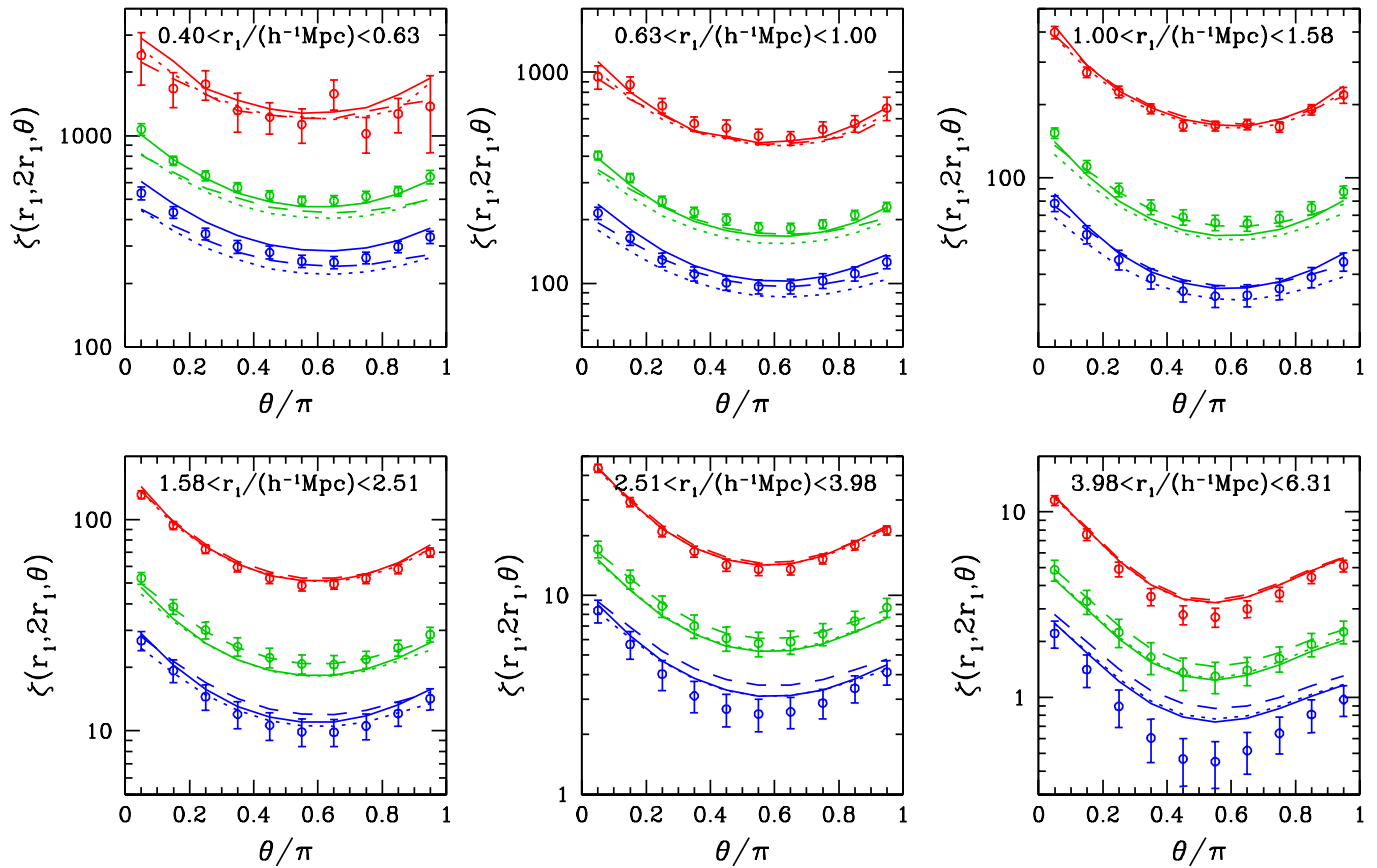}
\caption{Redshift-space 3PCFs with the triangle configuration of $r_2=2r_1$ for the three luminosity-threshold samples from scales of $r_1\sim 0.5\mpchi$ to $r_1 \sim 5\mpchi$. The 3PCF measurements of the $M_r<-19$, $M_r<-20$ and $M_r<-21$ are shown as the blue circles, green squares, and red triangles, respectively. 
Different lines display the predictions from three different models, the HOD model and the two SCAM models (using $V_{\rm acc}$ and $V_{\rm peak}$, respectively). All the models are obtained by fitting the projected and redshift-space 2PCFs (G16). For clarity, the measurements and predictions for the $M_r<-20$ and $M_r<-21$ samples are shifted upwards by 0.2 dex and 0.4 dex, respectively.
} \label{fig:zeta}
\end{figure*}

The 3PCF measurements are shown as the data points in Figure~\ref{fig:zeta}, with triangle configuration of $r_2=2r_1$ for the three luminosity-threshold samples (blue for $M_r<-19$, green for $M_r<-20$, and red for $M_r<-21$), from scales of $r_1=0.4\mpchi$ to $r_1=6.3\mpchi$. For a given triangle configuration, the amplitude of the 3PCFs is higher for more luminous galaxies. For a given sample, the amplitude of the 3PCFs increases toward smaller scales. For all the samples and for all the scales, the 3PCFs show clear dependence on the shape of galaxy triplets, with the amplitude increasing toward degenerate triangles ($\theta\sim 0$ or $\theta \sim \pi$). The overall trends of the scale and shape dependences in the 3PCFs agree with previous work \citep{Nichol06,Kulkarni07,Marin11,McBride11a,Guo14a,Moresco16}, but we are probing smaller scales than the previous literature, which can provide tighter constraints to the models as will be discussed in the following. We measure the 3PCFs from the HOD/SCAM mock catalogs and compare them to the measurements from the SDSS data. The model predictions are shown as curves in Figure~\ref{fig:zeta}, with solid lines for the HOD model, dotted lines for the $V_{\rm acc}$ model, and dashed lines for the $V_{\rm peak}$ model. 

To see how the three 2PCF-determined models work in predicting the 3PCFs, we compute and compare the $\chi^2/\rm{dof}$ values based on the predicted and 
measured 3PCFs, which are listed in Table~\ref{tab:chi2}, together with those for the 2PCFs in G16. In detail, the $\chi^2$ for the 3PCF is determined from
\begin{equation}
	\chi^2= \bm{(\zeta-\zeta^*)^T C^{-1} (\zeta-\zeta^*)},\label{eq:chi2}
\end{equation}
where $\bm{C}$ is the full error covariance matrix of the 3PCF data vector $\bm{\zeta}(r_1,2r_1,\theta)$. The quantity with (without) a superscript `$*$' is the one from the measurement (model). The covariance matrix is estimated using the jackknife resampling method with 400 subsamples as in G16. 

For the 2PCF measurements of each sample in G16, we have 12 bins for each of the 4 2PCF statistics (the projected 2PCFs and three redshift-space 2PCF multipoles) and one number density,   and each of the three models has 6 free parameters. The dof for the 2PCFs of each sample in G16 is then 43 ($\rm{dof}=4\times 12+1-6$).

For the 3PCF measurements, the 400 jackknife subsamples are not enough to accurately estimate the covariace matrices. We apply the singular value decomposition (SVD) to reduce the effect of noise in the covariance matrix, and only retain the dominant eigenmodes that have eigenvalues of $\lambda^2>\sqrt{2/N_{\rm jk}}$ \citep{Gaztanaga05}, where $N_{\rm jk}$ is the number of the jackknife subsamples. The number of removed eigenmodes, $N_{\rm SVD}$, is subtracted from the dof. As shown in Figure~\ref{fig:zeta}, we measure the 3PCF in 6 different $r_1$ bins and each bin has 10 data points at different values of $\theta$. In total, we have $N_\zeta=60$ data points of the 3PCFs for each sample. As we do not perform model fitting to the 3PCFs, we compute the 3PCF dof as ${\rm dof}=N_\zeta-N_{\rm SVD}$.

\begin{table}
\centering
\caption{Values of $\chi^2/\rm{dof}$ from 3PCF and 2PCF Measurements and Model Predictions} \label{tab:chi2}
\begin{tabular}{lrrr}
\hline
$\chi^2/\rm{dof}$ (3PCF)  &   HOD    & $V_{\rm acc}$ & $V_{\rm peak}$\\
\hline
$M_r<-19$  & $26.75/14$ & $32.43/14$ & $42.58/14$\\
$M_r<-20$  & $15.12/21$ & $33.64/21$ & $48.52/21$ \\
$M_r<-21$  & $47.88/44$ & $35.24/44$ & $37.89/44$ \\
\hline\hline
$\chi^2/\rm{dof}$ (2PCF)  &   HOD      & $V_{\rm acc}$ & $V_{\rm peak}$\\
\hline
$M_r<-19$  & $41.30/43$ & $56.71/43$ & $59.54/43$\\
$M_r<-20$  & $51.30/43$ & $64.51/43$ & $76.65/43$\\
$M_r<-21$  & $52.58/43$ & $46.43/43$ & $44.82/43$\\
\hline
\\
\end{tabular}
\tablecomments{The dof corresponding to the $\chi^2$ of 3PCFs is calculated as ${\rm dof}=N_{\zeta}-N_{\rm SVD}$, where $N_{\zeta}=60$ is the number of data points of $\zeta$ and $N_{\rm SVD}$ is the number of removed eigenmodes from the SVD analysis (see text). The $\chi^2/\rm{dof}$ for the bestfit models of 2PCFs in G16 are also shown in the bottom table for comparison.}
\end{table}
Generally, the HOD and SCAM models show different predictions of the 3PCFs on both small and intermediate scales. For the most luminous galaxy sample of $M_r<-21$, all the three models provide reasonable fits to the data, with $\chi^2/\rm{dof}$ varying from 0.80 to 1.09. As seen from Figure~\ref{fig:zeta}, except for the smallest scale (top-left panel), the three model predictions for this sample are comparable to each other. For the two fainter galaxy samples, the HOD model has much smaller $\chi^2$ values than the two SCAM models. The overall trend is similar to the 2PCF case --- for luminous samples (with luminosity threshold above $L^*$) all three models are reasonable, while for faint samples (with luminosity threshold below $L^*$) HOD model shows better performances. We note that the $2\sigma$ ranges of the expected $\chi^2$ distribution for the 3PCFs of the $M_r<-19$, $-20$ and $-21$ samples are approximately $14\pm10.6$, $21\pm13.0$, and $44\pm18.8$, respectively. For faint samples, the differences in the $\chi^2$ values of the three models are much more significant with the 3PCFs than those with the 2PCFs, implying that 3PCFs can help further discriminate different models. 

\begin{figure}
	\centering
	\includegraphics[width=0.49\textwidth]{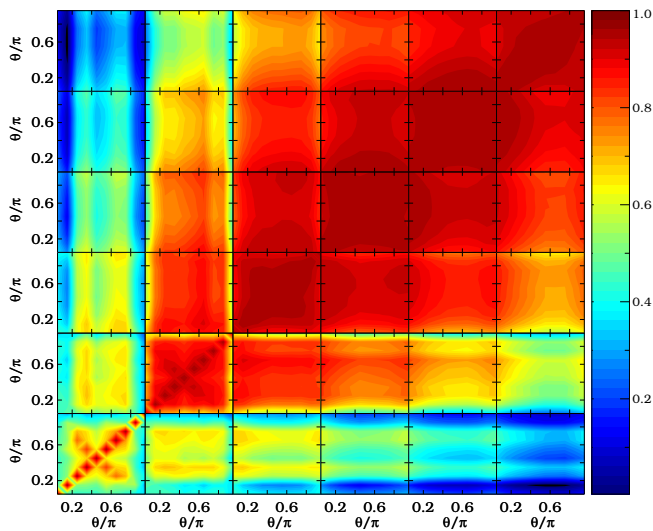}
	\caption{Normalized covariance matrix of the 3PCFs for the sample of $M_r<-20$. From left to right and bottom to top, the normalized covariance matrix is shown for the six $r_1$ bins from small to large scales, as labeled in Fig.~\ref{fig:zeta}.} \label{fig:cov3pcf}
\end{figure}
Figure~\ref{fig:zeta} suggests that the discrimination power has a large contribution from small scale 3PCFs. This is further supported by the structure in the covariance matrix. Figure~\ref{fig:cov3pcf} shows the normalized covariance matrix of the 3PCF measurements for the $M_r<-20$ sample. On large scales ($r_1\gtrsim 1\mpchi$), there is strong correlation between the different $\theta$ bins at each $r_1$ bin. On small scales, the correlation is approximately between $\theta$ and $\pi-\theta$ bins. This is in the Fingers-of-God regime, and such a correlation is expected from one-halo galaxy triplets (see below). On scales of $r_1>0.63\mpchi$, the measurements in different $r_1$ bins are also strongly correlated with each other (much stronger than those seen for the luminous sample in \citealt{Guo15b}). The differences among the different model predictions on large scales (being mainly an amplitude shift; see Fig.~\ref{fig:zeta}) therefore do not contribute significantly to the differences in the $\chi^2$ values. 

To further investigate the constraining power of the 3PCFs from different scales, we perform a test with the $M_r<-20$ sample by removing the 3PCF measurements in a $r_1$ bin at a time and compare the resulting $\chi^2/\rm{dof}$. We find that for all three models, the resulting $\chi^2/\rm{dof}$ is largest when the 3PCF measurements from the smallest $r_1$ bin ($0.4$--$0.63\mpchi$) are dropped. The small-scale 3PCFs of this sample provides the most discrimination power, consistent with the finding of \cite{Guo15a}. The HOD sample clearly shows a better fit to the small-scale data, which leads to the corresponding smallest $\chi^2$ value (15.1). Even though the $V_{\rm peak}$ model prediction agrees better (by eye) with the data points on large scales, it does not fit the smallest scale ($r_1<0.63\mpchi$) data and the overall $\chi^2$ (48.5) is much larger.

For the $M_r<-19$ sample, the HOD model also has the smallest $\chi^2$ value. The covariance matrix for the $M_r<-19$ sample is similar to that of the $M_r<-20$ sample with strong correlations shown on all scales. The small volume of the $M_r<-19$ sample leads to a noisy covariance matrix using the jackknife method, with 46 eigenmodes dropped from the SVD analysis, which reduces the constraining power of its 3PCF.

The three models determined from fitting the projected and redshift-space 2PCFs show differences in the redshift-space 3PCF predictions, suggesting that 3PCFs can help further constrain models. To quantify the improvements in model constraints by adding the 3PCFs, one needs to perform a joint fit to both 2PCFs and 3PCFs. In general, the bestfit model from the joint modeling will shift with respect to the 2PCF-determined model, and the relative contributions from the 2PCFs and 3PCFs will be different from what we have in this paper. While there are accurate and efficient methods to obtain model predictions for the 2PCFs \citep[e.g.,][]{Zheng16}, an accurate and efficient method for predicting the model 3PCFs is demanding. We therefore focus on models that reproduce the 2PCFs and investigate their performances for the 3PCFs to gain some general insights. There are efforts to perform joint modeling with 3PCFs calculated based on mock catalogs. For example, \cite{Guo15a} jointly model the 2PCFs and 3PCFs of $z\sim 0.5$ luminous red galaxies in the SDSS-III Baryon Oscillation Spectroscopic Survey \citep{Dawson13}. Indeed, they found that the inclusion of the 3PCF tightens the constraints on the HOD parameters, especially with the small-scale measurements. The implications from the study with the 2PCF-determined models here qualitatively agree with their results based on the joint 2PCF and 3PCF modeling.

The mock catalogs created in this work not only allow us to measure the model predictions but also enable us to decompose the model 3PCFs into various components. In what follows, we use the $M_r<-20$ sample as an example to show the differences in the 3PCF components from different models, which can enhance our understanding of the models.

\begin{figure}
	\centering
	\includegraphics[width=0.49\textwidth]{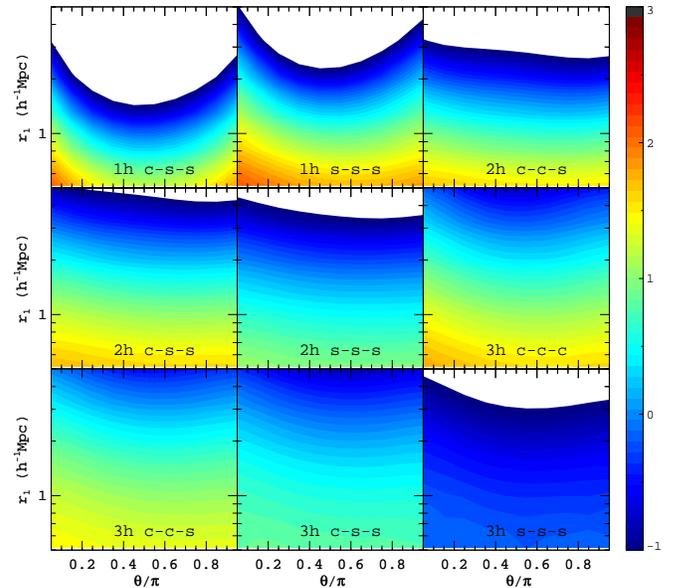}
	\caption{Decomposition of the total 3PCF $\zeta(r_1,2r_1,\theta)$ in Fig.~\ref{fig:zeta} into the contributions from the different halo components for the HOD model, where `c' stands for central galaxies and `s' for satellite galaxies. The labels `1h', `2h', and `3h' refer to the contributions from the one-halo, two-halo, and three-halo terms. For example, `1h c-s-s' means the contribution from the one-halo central-satellite-satellite galaxy triplets. The color contours are shown for $\log\zeta(r_1,2r_1,\theta)$, with redder colors for larger $\zeta$, and white area reflects the fact that there are no measured triplets in these regions.} \label{fig:decom_hod}
\end{figure}
In Figure~\ref{fig:decom_hod}, we show the decomposition in the HOD model for the total 3PCF of the $M_r<-20$ sample, as a function of scale and angle. The total 3PCF is decomposed into the contributions of one-halo, two-halo, and three-halo components (denoted as `1h', `2h' and `3h', respectively). The different halo components are further divided into contributions of galaxy triplets formed by different combinations of central (denoted as `c' in the figure) and satellite (denoted as `s') galaxies, as labeled. On small scales (a few tenth $\mpchi$), the one-halo term has a large contribution, as expected. In particular, it dominates the signal for degenerate triangle configurations ($\theta\sim 0$ or $\theta\sim \pi$), which results from the triplets elongated along the line of sight caused by the non-linear redshift-space distortion (the Fingers-of-God effect). On large scales (a few $\mpchi$), the three-halo term (especially from central-central-central triplets) dominates, with two-halo components involving central galaxies contributing substantially.

\begin{figure}
  \epsscale{1.15}
  \plotone{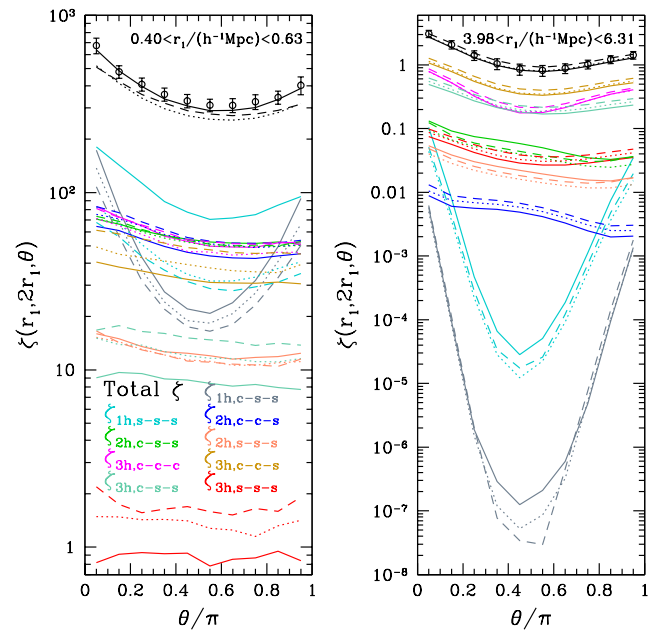}
  \caption[]{
  \label{fig:decomp_diff}
  Comparison of the 3PCF decompositions in different models. Solid, dotted, and dashed curves are for the HOD, $V_{\rm acc}$, and $V_{\rm peak}$ model, respectively. For each model, the different 3PCF components are represented by curves with different colors. Even though some components have almost negligible contributions to the total 3PCFs (e.g., the three-halo terms in the left panel and the one-halo terms in the right panel), we show all of them for completeness. 
  }
\end{figure}

The decompositions of the two SCAM models ($V_{\rm acc}$ and $V_{\rm peak}$) models have similar trends as for the HOD model. However, the three models are different in details. In Figure~\ref{fig:decomp_diff}, we compare the decompositions of the three models on small ($r_1\sim 0.5\mpchi$) and large ($r_1\sim 5\mpchi$) scales. On scales of $r_1<1\mpchi$, the HOD model agrees with the data better (Fig.~\ref{fig:zeta}). From Figure~\ref{fig:decomp_diff}, it is clear that
the differences in the models are mainly caused by the one-halo term. The HOD model has significantly larger contributions from the one-halo 
satellite-satellite-satellite triplets (`1h s-s-s'), while such triplet contributions in the two SCAM models are less substantial (lower than the HOD model by a factor of about two). As shown in Figures~14 and 15 of G16, in the SCAM models, satellite galaxies tend to populate slightly less massive halos and move faster (higher satellite velocity bias) than in the HOD model. Even though the SCAM models have a higher satellite fraction, the fast motion of satellites dilutes the amount of small-scale galaxy triplets, reducing the one-halo `c-s-s' and `s-s-s' triplets. We note that even at such small scales, the two-halo and three-halo terms still have significant contributions. This is a consequence of the redshift-space distortion effects that change the line-of-sight distribution of galaxies 
in redshift-space (see Fig.~2 of \citealt{Zheng16} for the case of the 2PCF, where the two-halo terms still have significant contributions to the small-scale Fingers-of-God feature). 

On large scales, the HOD model seems to agree with the $V_{\rm acc}$ model quite well. Figure~\ref{fig:decomp_diff} shows that the dominant contributions, the three-halo `c-c-c' and `c-c-s' triplets, have similar 3PCF amplitudes for the two models. There are differences in other components, e.g., in the HOD model higher contributions from the `1h c-s-s', `1h s-s-s', `2h c-s-s', and `2h s-s-s' components and smaller contributions from the `3h c-s-s' and `3h s-s-s' terms. The central galaxy occupation functions for these two models are quite similar (G16), as they are well constrained by the sample number density and the large-scale galaxy bias \citep{Zheng07,Zehavi11,Guo14b}. The above subtle differences mainly originate from the different satellite galaxy occupation functions and satellite 
velocity bias. These differences are trivial, as their contribution to the total 3PCFs is insignificant.

Comparing the different components of the $V_{\rm acc}$ and $V_{\rm peak}$ models, the $V_{\rm peak}$ model usually has higher amplitudes on scales of $r>1\mpchi$. For smaller scales, the $V_{\rm peak}$ model is only slightly lower for the two one-halo components of `1h c-s-s' and `1h s-s-s', and much larger for other components. This is consistent with what we find in Fig.~\ref{fig:zeta} and Table~\ref{tab:chi2}, where the $V_{\rm peak}$ model generally has a higher 3PCF amplitude than the $V_{\rm acc}$ model.

The decomposition also demonstrates that the shape dependence (i.e., dependence on $\theta$) of the 3PCF varies in the different components. The one-halo terms show the strongest dependence. The nonlinear redshift-space distortion (Fingers-of-God effect) makes the distribution of galaxies inside halos appear to be elongated along the line of sight, favoring triplets close to degenerate triangles ($\theta\sim0$ or $\theta\sim \pi$). The strong shape dependence of the one-halo terms dominates on small scales, but plays almost no role on large-scale 3PCFs, as their contribution to the total 3PCF diminishes. The large-scale shape dependence comes from the three-halo terms, reflecting the filamentary structure on linear or weakly nonlinear scales from gravitational evolution \citep{Scoccimarro01,Nichol06}. 

The radial distributions of the satellite galaxies in the HOD and SCAM models are different (G16), with the HOD model having a steeper satellite distribution. This difference (as well as the mean occupation function of satellites) contributes to the difference in the small-scale 3PCFs from the three models. The shape of the halo, defined by the satellite galaxies, can be another factor, as the relative mass scale shift among the three models corresponds to halo shape change. However, as already discussed, the shape dependence of the small-scale 3PCF mainly results from redshift-space distortion, the (real-space) shape of halos is not expected to play a big role. To test this, we compare the halo shape in the HOD and SCAM models. Given that halos can be well described by triaxial ellipsoids \citep{Jing02,Despali16,Vega16}, we follow the method laid out in \cite{Allgood06} to compute the eigenvectors of the reduced inertial tensor for the triaxial ellipsoids traced by satellite galaxies in the three models, and the directions of the eigenvectors define the three principal halo axes. We perform the calculation for all the halos with three or more satellites in the three models.

\begin{figure}
	\epsscale{1.15}\plotone{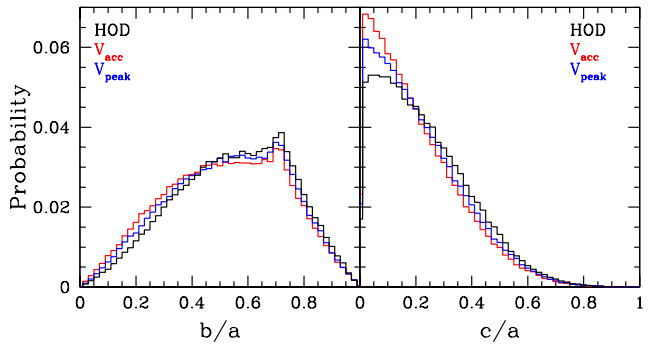}
	\caption{
		Comparison of the halo shape defined by the distribution of satellite galaxies
		in the three different models. The halo shape is characterized by the three
		semi-principal axes, $c\le b \le a$. Probability distributions of axis ratios, 
		$b/a$ and $c/a$ are shown in the left and right panel, respectively.} 
	\label{fig:shape}
\end{figure}
Figure~\ref{fig:shape} shows the probability distributions of the two axis ratios of $b/a$ and $c/a$ in the three models, assuming the lengths of the semi-principal axes of $c\le b\le a$. The overall distributions of the axis ratios in the three models are quite similar. The average axis ratios in all three models are around $b/a=0.5$ and $c/a=0.2$. The probability distributions of the axis ratios roughly keep the same even if we consider the satellite galaxy distributions within different distances to the halo centers. While the overall halo shape distribution marginalized over the full model parameter spaces could be different for the case shown here, the comparison made in Figure~\ref{fig:shape} is to check the effect of the halo shape on the bestfit models shown in Fig.~\ref{fig:zeta}.
The comparison implies that the differences seen in the model predictions (Fig.~\ref{fig:zeta}) are the result of the differences in the radial distribution, the mean occupation function, and the velocity bias of satellite galaxies, rather than the shape distribution of satellite galaxies. 
In addition, given the large role played by the small-scale redshift-space distortion in the shape dependence of the small-scale 3PCFs, it is unlikely 
that the intrinsic halo shape (defined by the satellite distribution) can be well constrained by the redshift-space 3PCFs.  

\section{Conclusion} \label{sec:discussion}

In this paper, we present the luminosity-dependent redshift-space galaxy 3PCF measurements for the SDSS DR7 Main galaxy sample. The measurements are then used to test three models, the HOD model and two SCAM models (based on $V_{\rm acc}$ and $V_{\rm peak}$). These models are obtained by fitting the projected and redshift-space galaxy 2PCFs (G16), and the 3PCF predictions come from mock catalogs.

For the most luminous galaxy sample ($M_r<-21$), the predictions from all the three models agree well with the 3PCF measurements. For the fainter samples 
($M_r<-20$ and $M_r<-19$), the prediction from the HOD model generally provides a better match to the measurements. This strengthens the trend seen in modeling galaxy 2PCFs (G16) -- all models work reasonably well for samples with luminosity threshold above $L^*$, while the HOD model stands out in better fitting faint samples (with luminosity threshold below $L^*$). 

We decompose the total 3PCFs into contributions from different types of galaxy triplets. On small scales, the 3PCFs are dominated by contributions from one-halo galaxy triplets, and the dependence on the triangle shape is driven by the redshift-space distortion (the Fingers-of-God effect). This is consistent with the fact that the 3PCFs in projected space have a much weaker shape dependence than those in redshift space, as found by \cite{Guo14a}. The intrinsic halo shape defined by the distribution of satellites, however, does not contribute much to the shape dependence of the 3PCFs, as the halo shape in redshift space is largely determined by the motion of satellites. On large scales, the 3PCFs are dominated by contributions from three-halo galaxy triplets, and the shape dependence reflects the filamentary structure from gravitational evolution. The small-scale 3PCFs, where the three model predictions more strongly deviate from each other (in both 3PCF amplitude and shape), are related to the radial distribution, the mean occupation function, and the velocity bias of satellite galaxies, so we expect that the 
small-scale 3PCFs can further constrain the above galaxy-halo connection and distinguish different models. 

In conclusion, besides the traditional application of the galaxy 3PCF to measure the nonlinear galaxy bias \citep[e.g.,][]{Guo09b,McBride11b,Marin13}, the additional information encoded in the 3PCF can be used to distinguish the various halo (and subhalo) models, as well as tighten the constraints on the occupation distribution of galaxies inside halos \citep[e.g.,][]{Kulkarni07,Guo15b}. More insights can be yielded by jointly modeling 2PCFs and 3PCFs in future work.

\section*{Acknowledgements}
We thank the anonymous referee for the helpful suggestions that improve the presentation of this paper. This work is supported by the 973 Program (No. 2015CB857003). HG acknowledges the support of the NSFC-11543003 and the 100 Talents Program of the Chinese Academy of Sciences. ZZ was partially supported by NSF grant AST-1208891 and NASA grant NNX14AC89G. PB was supported through program number HST-HF2-51353.001-A, provided by NASA through a Hubble Fellowship grant from STScI, which is operated by the Association of Universities for Research in Astronomy, Incorporated, under NASA contract NAS5-26555. IZ acknowledges support by NSF grant AST-1612085 and by a CWRU Faculty Seed Grant. JC, GF, SG, AK, FP and SRT acknowledge support from the Spanish MICINNs Consolider-Ingenio 2010 Programme under grant MultiDark CSD2009-00064, MINECO Centro de Excelencia Severo Ochoa Programme under grant SEV-2012-0249, and MINECO grant AYA2014-60641-C2-1-P. GY acknowledges financial support from MINECO (Spain) under research grants AYA2012-31101 and FPA2012-34694.

We gratefully acknowledge the use of the High Performance Computing Resource in the Core Facility for Advanced Research Computing at Shanghai Astronomical Observatory, the Gauss Centre for Supercomputing e.V. (www.gauss-centre.eu) and the Partnership for Advanced Supercomputing in Europe (PRACE, www.prace-ri.eu) for funding the MultiDark simulation project by providing computing time on the GCS Supercomputer SuperMUC at Leibniz Supercomputing Centre (LRZ, www.lrz.de). The CosmoSim database (www.cosmosim.org) used in this paper is a service by the Leibniz-Institute for Astrophysics Potsdam (AIP). The MultiDark database was developed in cooperation with the Spanish MultiDark Consolider Project CSD2009-00064.


\begin{thebibliography}

	\bibitem[\protect\citeauthoryear{{Abazajian} et~al.,}{{Abazajian} et~al.}{2009}]{Abazajian09}
	{Abazajian} K.~N.  et~al., 2009, \apjs, 182, 543
	
	\bibitem[{Allgood} et al.(2006)]{Allgood06}
	{Allgood}, B., {Flores}, R.~A., {Primack}, J.~R., et al. 2006, \mnras, 367,
	1781
	
	\bibitem[{Behroozi} et al.(2010)]{Behroozi10}
	{Behroozi}, P.~S., {Conroy}, C., \& {Wechsler}, R.~H. 2010, \apj, 717, 379
	
	\bibitem[{Behroozi} et al.(2013)]{Behroozi13}
	{Behroozi}, P.~S., {Wechsler}, R.~H., \& {Wu}, H.-Y. 2013, \apj, 762, 109
	
	\bibitem[{Berlind} \& {Weinberg}(2002)]{Berlind02}
	{Berlind}, A.~A., \& {Weinberg}, D.~H. 2002, \apj, 575, 587
	
	\bibitem[{Bernardeau} et al.(2002)]{Bernardeau02}
	{Bernardeau}, F., {Colombi}, S., {Gazta{\~n}aga}, E., \& {Scoccimarro}, R.
	2002, \physrep, 367, 1
	
	\bibitem[{Conroy} et al.(2006)]{Conroy06}
	{Conroy}, C., {Wechsler}, R.~H., \& {Kravtsov}, A.~V. 2006, \apj, 647, 201

	\bibitem[{Dawson} et al.(2013)]{Dawson13}
	{Dawson} K.~S.  et~al., 2013, \aj, 145, 10
	
	\bibitem[Despali et al.(2016)]{Despali16} 
	Despali, G., Giocoli, C., Bonamigo, M., Limousin, M., \& Tormen, G.\ 2016, arXiv:1605.04319
	
	\bibitem[{Gao} et al.(2004)]{Gao04}
	{Gao}, L., {De Lucia}, G., {White}, S.~D.~M., \& {Jenkins}, A. 2004, \mnras,
	352, L1
	
	\bibitem[{Gazta{\~n}aga} et al.(2009)]{Gaztanaga09}
	{Gazta{\~n}aga}, E., {Cabr{\'e}}, A., {Castander}, F., {Crocce}, M., \&
	{Fosalba}, P. 2009, \mnras, 399, 801
	
	\bibitem[{Gazta{\~n}aga} \& {Scoccimarro}(2005)]{Gaztanaga05}
	{Gazta{\~n}aga}, E., \& {Scoccimarro}, R. 2005, \mnras, 361, 824
	
	\bibitem[{Gaztanaga} \& {Frieman}(1994)]{Gaztanaga94}
	{Gaztanaga}, E., \& {Frieman}, J.~A. 1994, \apjl, 437, L13
	
	\bibitem[{Guo} \& {Jing}(2009)]{Guo09b}
	{Guo}, H., \& {Jing}, Y.~P. 2009, \apj, 702, 425
	
	\bibitem[{Guo} et al.(2014a)]{Guo14a}
	{Guo}, H., {Li}, C., {Jing}, Y.~P., \& {B{\"o}rner}, G. 2014a, \apj, 780, 139
	
	\bibitem[{Guo} et al.(2014b)]{Guo14b}
	{Guo}, H., {Zheng}, Z., {Zehavi}, I., et al. 2014b, \mnras, 441, 2398
	
    \bibitem[{Guo} et al.(2016)]{Guo16} 
    {Guo}, H., {Zheng}, Z., {Behroozi}, P. S., et al., 2016, \mnras, 459, 3040 (G16)
	
	\bibitem[{Guo} et al.(2015a)]{Guo15a}
	{Guo}, H., {Zheng}, Z., {Jing}, Y.~P., et al. 2015a, \mnras, 449, L95
	
	\bibitem[{Guo} et al.(2015b)]{Guo15b}
	{Guo}, H., {Zheng}, Z., {Zehavi}, I., et al. 2015b, \mnras, 453, 4368
	
	\bibitem[{Guo} et al.(2010)]{Guo10}
	{Guo}, Q., {White}, S., {Li}, C., \& {Boylan-Kolchin}, M. 2010, \mnras, 404,
	1111
	
	\bibitem[{Jing} \& {B{\"o}rner}(2004)]{Jing04}
	{Jing}, Y.~P., \& {B{\"o}rner}, G. 2004, \apj, 607, 140
	
	\bibitem[{Jing} et al.(1998)]{Jing98}
	{Jing}, Y.~P., {Mo}, H.~J., \& {B{\"o}rner}, G. 1998, \apj, 494, 1
	
	\bibitem[{Jing} \& {Suto}(2002)]{Jing02}
	{Jing}, Y.~P., \& {Suto}, Y. 2002, \apj, 574, 538
	
	\bibitem[Klypin et al.(2016)]{Klypin16} 
	Klypin, A., Yepes, G., Gottl{\"o}ber, S., Prada, F., \& He{\ss}, S.\ 2016, \mnras, 457, 4340 
	
	\bibitem[{Kravtsov} et al.(2004)]{Kravtsov04}
	{Kravtsov}, A.~V., {Berlind}, A.~A., {Wechsler}, R.~H., et al. 2004, \apj, 609,
	35
	
	\bibitem[{Kulkarni} et al.(2007)]{Kulkarni07}
	{Kulkarni}, G.~V., {Nichol}, R.~C., {Sheth}, R.~K., et al. 2007, \mnras, 378,
	1196
	
    \bibitem[Mar{\'{\i}}n(2011)]{Marin11} 
    Mar{\'{\i}}n, F.\ 2011, \apj, 737, 97 	
	
	\bibitem[{Mar{\'{\i}}n} et al.(2013)]{Marin13}
	{Mar{\'{\i}}n}, F.~A., {Blake}, C., {Poole}, G.~B., et al. 2013, \mnras, 432,
	2654
	
    \bibitem[McBride et al.(2011)]{McBride11a} 
    McBride, C.~K., Connolly, A.~J., Gardner, J.~P., et al.\ 2011, \apj, 726, 13 	
	
	\bibitem[{McBride} et al.(2011)]{McBride11b}
	{McBride}, C.~K., {Connolly}, A.~J., {Gardner}, J.~P., et al. 2011, \apj, 739,
	85
	
	\bibitem[Moresco et al.(2014)]{Moresco14} 
	Moresco, M., Marulli, F., Baldi, M., Moscardini, L., \& Cimatti, A.\ 2014, \mnras, 443, 2874	
	
	\bibitem[Moresco et al.(2016)]{Moresco16} 
	Moresco, M., Marulli, F., Moscardini, L., et al.\ 2016, arXiv:1603.08924 
	
	\bibitem[{Moster} et al.(2010)]{Moster10}
	{Moster}, B.~P., {Somerville}, R.~S., {Maulbetsch}, C., et al. 2010, \apj, 710,
	903
	
	\bibitem[{Nichol} et al.(2006)]{Nichol06}
	{Nichol}, R.~C., {Sheth}, R.~K., {Suto}, Y., et al. 2006, \mnras, 368, 1507
	
	\bibitem[{Nuza} et al.(2013)]{Nuza13}
	{Nuza}, S.~E., {S{\'a}nchez}, A.~G., {Prada}, F., et al. 2013, \mnras, 432, 743
	
	\bibitem[{Pan} \& {Szapudi}(2005)]{Pan05}
	{Pan}, J., \& {Szapudi}, I. 2005, \mnras, 362, 1363
	
	\bibitem[{Peacock} \& {Smith}(2000)]{Peacock00}
	{Peacock}, J.~A., \& {Smith}, R.~E. 2000, \mnras, 318, 1144
	
	\bibitem[{Pujol} et al.(2014)]{Pujol14}
	{Pujol}, A., {Gazta{\~n}aga}, E., {Giocoli}, C., et al. 2014, \mnras, 438, 3205
	
	\bibitem[{Rodr{\'{\i}}guez-Puebla} et al.(2013)]{Rodriguez-Puebla13}
	{Rodr{\'{\i}}guez-Puebla}, A., {Avila-Reese}, V., \& {Drory}, N. 2013, \apj,
	767, 92
	
	\bibitem[{Sawala} et al.(2015)]{Sawala15}
	{Sawala}, T., {Frenk}, C.~S., {Fattahi}, A., et al. 2015, \mnras, 448, 2941
	
	\bibitem[{Scoccimarro} et al.(2001)]{Scoccimarro01}
	{Scoccimarro}, R., {Sheth}, R.~K., {Hui}, L., \& {Jain}, B. 2001, \apj, 546, 20
	
	\bibitem[Sefusatti \& Scoccimarro(2005)]{Sefusatti05} 
        Sefusatti, E., \& Scoccimarro, R.\ 2005, \prd, 71, 063001
	
	\bibitem[{Slepian} \& {Eisenstein}(2015)]{Slepian15a}
	{Slepian}, Z., \& {Eisenstein}, D.~J. 2015, \mnras, 448, 9
	
	\bibitem[{Smith} et al.(2008)]{Smith08}
	{Smith}, R.~E., {Sheth}, R.~K., \& {Scoccimarro}, R. 2008, \prd, 78, 023523
	
	\bibitem[{Szapudi} \& {Szalay}(1998)]{Szapudi98}
	{Szapudi}, I., \& {Szalay}, A.~S. 1998, \apjl, 494, L41
	
	\bibitem[{Vale} \& {Ostriker}(2006)]{Vale06}
	{Vale}, A., \& {Ostriker}, J.~P. 2006, \mnras, 371, 1173
	
	\bibitem[Vega et al.(2016)]{Vega16} 
	Vega, J., Yepes, G., \& Gottl{\"o}ber, S.\ 2016, arXiv:1603.02256 
	
	\bibitem[{Wang} et al.(2007)]{Wang07}
	{Wang}, Y., {Yang}, X., {Mo}, H.~J., \& {van den Bosch}, F.~C. 2007, \apj, 664,
	608
	
	\bibitem[\protect\citeauthoryear{Wang et al.}{2014}]{Wang14}
	Wang W., Sales L.~V., Henriques B.~M.~B., White S.~D.~M., 2014, \mnras, 442, 1363

	\bibitem[\protect\citeauthoryear{{York} et~al.,}{{York} et~al.}{2000}]{York00} 
	{York} D.~G.  et~al., 2000, \aj, 120, 1579
	
	\bibitem[{Zehavi} et al.(2011)]{Zehavi11}
	{Zehavi}, I., {Zheng}, Z., {Weinberg}, D.~H., et al. 2011, \apj, 736, 59
	
	\bibitem[{Zheng}(2004)]{Zheng04}
	{Zheng}, Z. 2004, \apj, 610, 61
	
	\bibitem[{Zheng} et al.(2007)]{Zheng07}
	{Zheng}, Z., {Coil}, A.~L., \& {Zehavi}, I. 2007, \apj, 667, 760

	\bibitem[{Zheng} \& {Guo}(2016)]{Zheng16} 
	{Zheng}, Z., \& {Guo}, H.\ 2016, \mnras, 458, 4015 
	
	\bibitem[{Zheng} et al.(2009)]{Zheng09}
	{Zheng}, Z., {Zehavi}, I., {Eisenstein}, D.~J., {Weinberg}, D.~H., \& {Jing},
	Y.~P. 2009, \apj, 707, 554
	
	\bibitem[{Zheng} et al.(2005)]{Zheng05}
	{Zheng}, Z., {Berlind}, A.~A., {Weinberg}, D.~H., et al. 2005, \apj, 633, 791
	
\end{thebibliography}
\end{document}